\documentclass{article}
\usepackage{spconf,amsmath,graphicx}

\usepackage{enumitem}
\setlist{nosep, leftmargin=14pt}

\usepackage{mwe} 

\usepackage{amsmath,amssymb,amsthm}
\usepackage{graphicx}
\graphicspath{figure/}
\usepackage{cite}
\usepackage{amsmath,amssymb,amsfonts}
\usepackage{algorithmic}
\usepackage{array}

\usepackage{graphicx}
\usepackage{textcomp}
\usepackage{lineno}

\usepackage{makecell}
\usepackage{stfloats}
\usepackage{float}
\usepackage{hhline}
\usepackage{multirow}
\usepackage{booktabs}

\title{Real-Time Limited-View CT Inpainting and Reconstruction with Dual Domain Based on Spatial Information}
\name{Ken Deng, Chang Sun, Yitong Liu, Hongwen Yang 
\thanks{K. Deng ,C. Sun, Y. Liu and H. Yang is with the Institute of Wireless Theories and Technologies Lab, Beijing University of Posts and Telecommunications, Haidian, Beijing 100876, China (e-mail: arieldeng@bupt.edu.cn; sc1998@bupt.edu.cn; liuyitong@bupt.edu.cn; yanghong@bupt.edu.cn).}
\thanks{Y. Liu is the corresponding author (e-mail: liuyitong@bupt.edu.cn).}}

\address{Beijing University of Posts and Telecommunications, Beijng 100089, China
	}
\begin{document}
	\bibliographystyle{IEEEtran}
	
	\title{Real-Time Limited-View CT Inpainting and Reconstruction with Dual Domain Based on Spatial Information}

	\maketitle
	
	\begin{abstract}
		Low-dose Computed Tomography is a common issue in reality. Current reduction, sparse sampling and limited-view scanning can all cause it. Between them, limited-view CT is general in the industry due to inevitable mechanical and physical limitation. However, limited-view CT can cause serious imaging problem on account of its massive information loss. Thus, we should effectively utilize the scant prior information to perform completion. It is an undeniable fact that CT imaging slices are extremely dense, which leads to high continuity between successive images. We realized that fully exploit the spatial correlation between consecutive frames can significantly improve restoration results in video inpainting. Inspired by this, we propose a deep learning-based three-stage algorithm that hoist limited-view CT imaging quality based on spatial information. In stage one, to better utilize prior information in the Radon domain, we design an adversarial autoencoder to complement the Radon data. In the second stage, a model is built to perform inpainting based on spatial continuity in the image domain. At this point, we have roughly restored the imaging, while its texture still needs to be finely repaired. Hence, we propose a model to accurately restore the image in stage three, and finally achieve an ideal inpainting result. In addition, we adopt FBP instead of SART-TV to make our algorithm more suitable for real-time use. In the experiment, we restore and reconstruct the Radon data that has been cut the rear one-third part, they achieve PSNR of 40.209, SSIM of 0.943, while precisely present the texture.
	\end{abstract}

	
	\section{Introduction}
	\label{sec:introduction}
	Computed Tomography has been successfully applied in medicine, biology, industry and other fields, providing huge help for industrial production, medical research and people's daily life \cite{chen2017low}. Nevertheless, the radiation dose brought by CT scanning may somehow have a negative effect on human body that cannot be neglect. Thus, it is crucial for CT scanning to lower its radiation dose \cite{kalra2004strategies} in accordance with ALARA (as low as reasonably achievable) \cite{slovis2002alara}. Low-dose Computed Tomography (LDCT) can be realized through current reduction, sparse sampling and limited-view scanning. Among these, limited-view CT is really general because that we often encounter mechanical and physical restriction in the industry which makes it difficult for the machine to scan through an object. Despite the general application of limited-view CT, its imaging leads to some grievous problems like blur \cite{khare2005a}, artifacts \cite{xie2018artifact,zhang2019dualres,xie2019artifact,zhang2020artifact} and low signal-to-noise ratio \cite{xie2012image,chen2017low}, they undoubtedly have a great influence on clinical diagnosis. Thus, it is crucial for researchers to fully utilize the limited prior information to effectively complement the fragmentary data. 
	
\begin{figure}[htbp]
		\centerline{\includegraphics[width=\columnwidth]{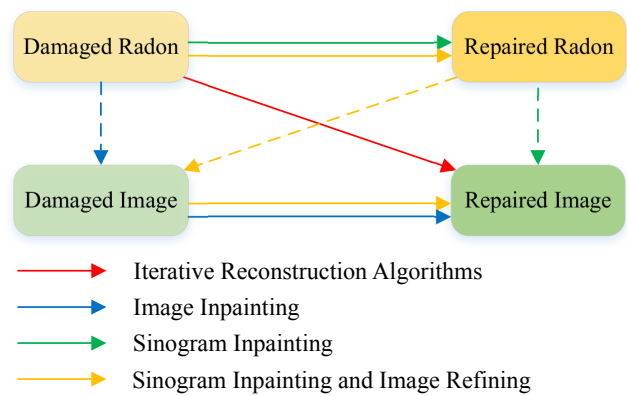}}
		\caption{The technology roadmap of prevailing CT inpainting and reconstruction algorithms, the dash line in this figure refers to the reconstruction step from the Radon domain to the image domain through FBP or SART-TV.}
		\label{sec:fig2-1}
		
	\end{figure}

	Traditional analytical reconstruction algorithms, such as FBP \cite{katsevich2002theoretically}, have high requirements for data integrifty. When the radiation dose is reduced, artifacts in reconstructed images will increase rapidly \cite{imai2009statistical}. In order to upgrade the quality of reconstructed images, many researchers have proposed various algorithms for LDCT imaging reconstruction, and we conclude them into several paths that are presented in Fig.\ref{sec:fig2-1} for better comprehension.
	
	\textbf{Iterative Reconstruction Algorithms} are represented by the red line in Fig.\ref{sec:fig2-1}, which can directly reconstruct damaged Radon data into target results in the image domain. Model-based iterative reconstruction (MBIR) algorithm \cite{liu2014model}, also known as statistical image reconstruction (SIR) method, combines the modeling of some key parameters to perform high-quality reconstruction of LDCT. Using image priors in MBIR can effectively improve the image reconstruction quality of LDCT scans \cite{chen2008prior,chen2011time}, while still have the high computational complexity.
	
	In addition to the prior information, various regularization methods have played a crucial role in iterative algorithms of CT reconstruction. The most typical regularization method is the total variation (TV) method \cite{rudin1992nonlinear}. In the light of TV, researchers came up with more reconstruction methods, such as TV-POCS \cite{sidky2008image}, TGV \cite{niu2014sparse} and SART-TV \cite{sidky2006accurate} which was proposed on the basis of SART \cite{andersen1984simultaneous}. Those algorithms can suppress image artifacts to a certain extent so as to improve imaging quality. In addition, dictionary learning is often used as a regularizer in MBIR algorithms \cite{xu2012low,cao2013limited,zhang2017low,xu2019l0dl}, and multiple dictionaries are beneficial to reducing artifacts caused by limited-view CT reconstruction.
	
	With the development of computing power, deep learning-based methods \cite{lecun2015deep,he2016deep,srivastava2015training,xie2012image,dong2015image,wurfl2016deep,mao2016image} have been applied to the restoration of LDCT reconstructed images in recent years. The methods can be roughly divided into the below three categories.
	
	\textbf{Image Inpainting} algorithms are presented by blue lines in Fig.1, they firstly reconstruct the damaged Radon data into the damaged image with artifacts, then reduce the artifacts and noises in the image domain. Lots of researchers are currently using convolutional neural network (CNN) and deep learning architecture to perform this procedure \cite{zhang2016image,chen2017low,kang2017deep,zhang2018sparse,xie2018artifact,wang2020deep,ronneberger2015u,zhang2019dualres,goodfellow2014generative,xie2019artifact,anirudh2019improving}. Zhang et al \cite{zhang2016image} proposed a data-driven learning method based on deep CNN. RED-CNN \cite{chen2017low} combines the autoencoder, deconvolutional network and shortcut connections into the residual encoder-decoder CNN for LDCT imaging. Kang et al \cite{kang2017deep} applied deep CNN to the wavelet transform coefficients of LDCT images, used directional wavelet transform to extract the directional component of artifacts. Wang et al \cite{wang2020deep} developed a limited-angle translational CT (TCT) image reconstruction algorithm based on U-Net \cite{ronneberger2015u}. Since Goodfellow et al. proposed Generative Adversarial Nets (GAN) \cite{goodfellow2014generative} in 2014, GAN has been widely used in various image processing tasks, including the post-processing of CT images. Xie et al. \cite{xie2019artifact} proposed an end-to-end conditional GAN with joint loss function, which can effectively remove artifacts.
	\begin{figure*}[tbp]
		\centering
		\includegraphics[scale=0.4]{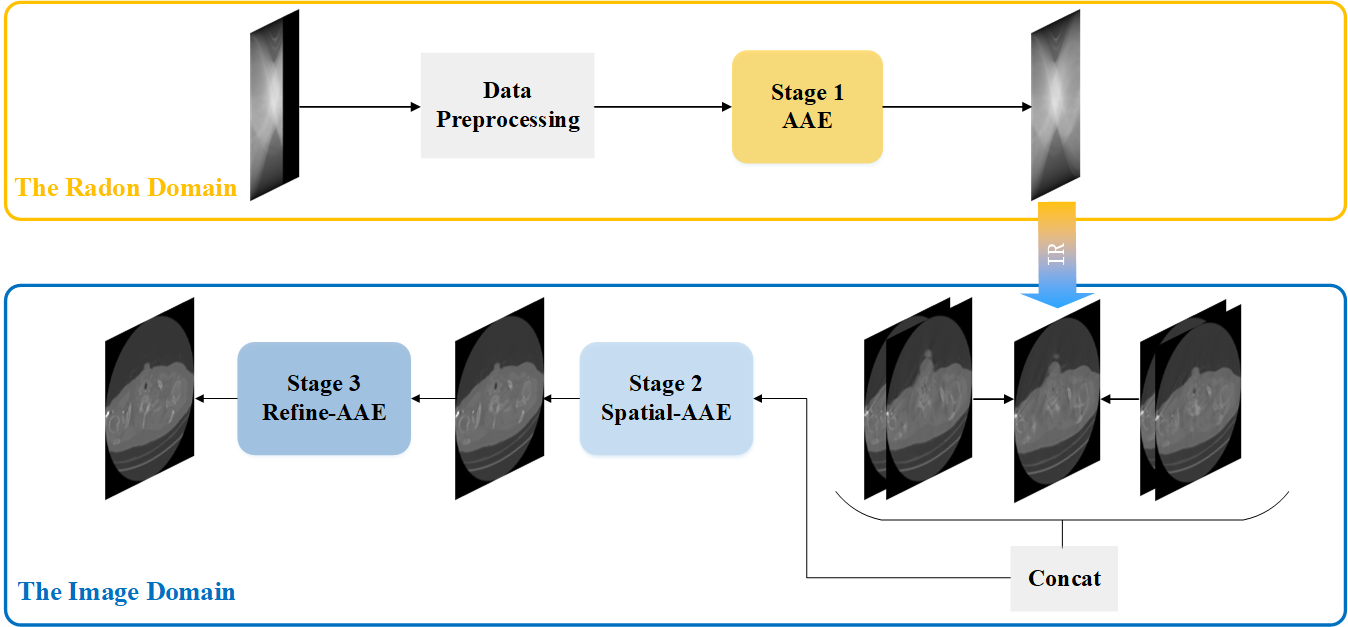}
		\caption{The overall architecture of our proposed three-stage restoration and reconstruction algorithm for limited-view CT imaging.}
		\label{sec:fig2-2}
		\vspace{-15pt}
	\end{figure*}
	
	\textbf{Sinogram Inpainting} algorithms are presented by green lines in Fig.\ref{sec:fig2-1}, they firstly restore the missing part in the Radon domain, then reconstruct it into the image domain to get the final result \cite{2019A,Li2019Promising,dai2018limited,anirudh2018lose,xiubin2016limited}. Li et al. \cite{2019A} proposed an effective GAN-based repairing method named patch-GAN, which trains the network to learn the data distribution of the sinogram to restore the missing sinogram data. In another paper \cite{Li2019Promising}, Li et al. proposed SI-GAN on the basis of \cite{zhang2018sparse}, using a joint loss function combining the Radon domain and the image domain to repair ``ultra-limited-angle" sinogram. In 2019, Dai et al. \cite{dai2018limited} proposed a limited-view cone-beam CT reconstruction algorithm. It slices the cone-beam projection data into the sequence of two-dimensional images, uses an autoencoder network to estimate the missing part, then stack them in order and finally use FDK \cite{feldkamp1984practical} for three-dimensional reconstruction. Anirudh et al. \cite{anirudh2018lose} transformed the missing sinogram into a latent space through a fully convolutional one-dimensional CNN, then used GAN to complement the missing part. Dai et al. \cite{xiubin2016limited} calculated the geometric image moment based on the projection-geometric moment transformation of the known Radon data, then estimated the projection-geometric moment transformation of the unknown Radon data based on the geometric image moment.
	
	\textbf{Sinogram Inpainting and Image Refining} algorithms are presented by yellow lines in Fig.\ref{sec:fig2-1}, they firstly restore the missing part in the Radon domain, then reconstruct the full-view Radon data into the image domain so as to finely repair the image to obtain higher quality \cite{hammernik2017deep,zhao2018unsupervised,zhao2018sparse,lee2019high,zhang2020artifact}. In 2017, Hammernik et al. \cite{hammernik2017deep} proposed a two-stage deep learning architecture, they first learn the compensation weights that account for the missing data in the projection domain, then they formulate the image restoration problem as a variational network to eliminate coherent streaking artifacts. Zhao et al. \cite{zhao2018unsupervised} proposed a GAN-based sinogram inpainting network, which achieved unsupervised training in a sinogram-image-sinogram closed loop. Zhao et al. \cite{zhao2018sparse} also proposed a two-stage method, firstly they use an interpolating convolutional network to obtain the full-view projection data, then use GAN to output high-quality CT images. In 2019, Lee et al. \cite{lee2019high} proposed a deep learning model based on fully convolutional network and wavelet transform. In the latest research, Zhang et al. \cite{zhang2020artifact} proposed an end-to-end hybrid domain CNN (hdNet), which consists of a CNN operating in the sinogram domain, a domain transformation operation, and a CNN operating in the image domain.
	
	Inspired by the combination of the two stages, we implement Radon data completion through our proposed adversarial autoencoder (AAE) in stage one. In the second and third stage, after enriching the information through Radon data completion, we construct the Radon data into the image domain and realize the image inpainting in a "coarse-to-fine" \cite{Zhou_2013_ICCV_Workshops} manner.
	
	However, all of the above algorithms merely focus on a single image slice while neglecting the abundant spatial correlation between consecutive image slices. Consequently, these algorithms may still have trouble to reach an ideal level of limited-view CT inpainting and reconstruction that can precisely presents the image texture.
	During our investigation of video inpainting \cite{claus2019videnn,tassano2020fastdvdnet}, we realize the significance of making full use of spatial correlation and continuity between consecutive image slices. Therefore, we propose an origin cascade model in stage two called Spatial-AAE to fully utilize the spatial continuity, thereby breaking the limitation of two-dimensional space. 
	
	It is also worth mentioning that, unlike other current limited-view CT inpainting and reconstruction algorithms, we use FBP \cite{katsevich2002theoretically} instead of SART-TV \cite{sidky2006accurate} to speed up the reconstruction process. Besides, our models do not limit resolution of the input data, therefore can be well generalized to various datasets. In our experiments, we compare our algorithm with the other four prevalent algorithms under four sorts of damaged data, exhibiting its prominent performance and robustness.
	
	The organization of this paper is as follows, Sec II presents the design details of our proposed algorithm and models, Sec III shows our experimental results, and we finally conclude our research work in Sec IV.
	
	
	\section{Methods}
	\label{sec:methods}
	This paper proposes a three-stage restoration and reconstruction algorithm for limited-view CT imaging, and its overall architecture is shown in Fig.\ref{sec:fig2-2}. In the first stage, after the limited-view Radon data is preprocessed, we input it into the Adversarial Autoencoder we designed for data completion to obtain the full-view Radon data. In the second stage, the output of stage one is first reconstructed into the image domain, and combined with two consecutive slices before and after to form a group, then we sent this group into our proposed Spatial-AAE model to perform image restoration based on spatial information. It is worth noting that through the above two stages of restoration and reconstruction, most of the texture in the image ground truth can be restored, but the result still cannot clearly reflect the precise details, which may pose obstacles for the practical applications. Therefore, we built the Refine-AAE high-precision inpainting network in stage three, utilizing the idea of "coarse-to-fine" \cite{Zhou_2013_ICCV_Workshops} in deep learning to refine the image in patches. The details of our algorithm are shown below.

	\subsection{Data Preprocessing}
	\label{rec:3.1}
\begin{figure}[htbp]
		\vspace{-10pt}
		\centerline{\includegraphics[scale=0.4]{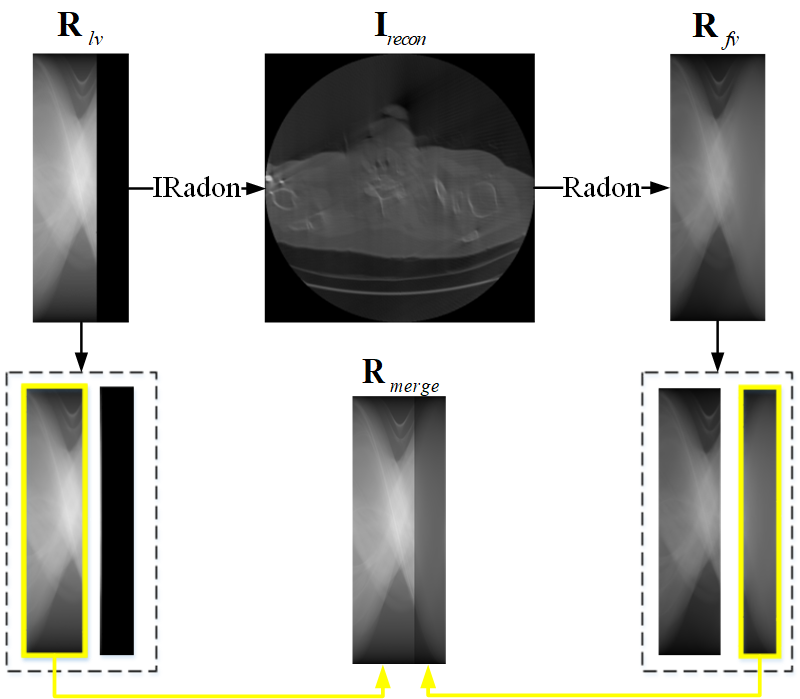}}
		\caption{Procedure of data preprocessing.}
		\label{sec:fig3-1}
		\vspace{-15pt}
	\end{figure}

	In order to provide more prior information, we adopt the data preprocessing method from paper \cite{xie2018artifact}, as shown in Fig.\ref{sec:fig3-1}. For the limited-view Radon data $\boldsymbol{R}_{lv}$, we first transform it into the image data $\boldsymbol{I}_{recon}$ through inverse radon transformation, and then convert the image into the full-view Radon data $\boldsymbol{R}_{fv}$ through Radon transformation. We crop this full-view Radon data for preliminary completion of the missing part in the original data, so as to obtain the fused Radon data $\boldsymbol{R}_{merge}$.
	
	\subsection{Algorithm Pipeline}
	\label{rec:3.2}
	
	\subsubsection{Stage 1: Limited-view Data Completion in the Radon Domain}
\begin{figure}[htbp]
		\centerline{\includegraphics[width=\columnwidth]{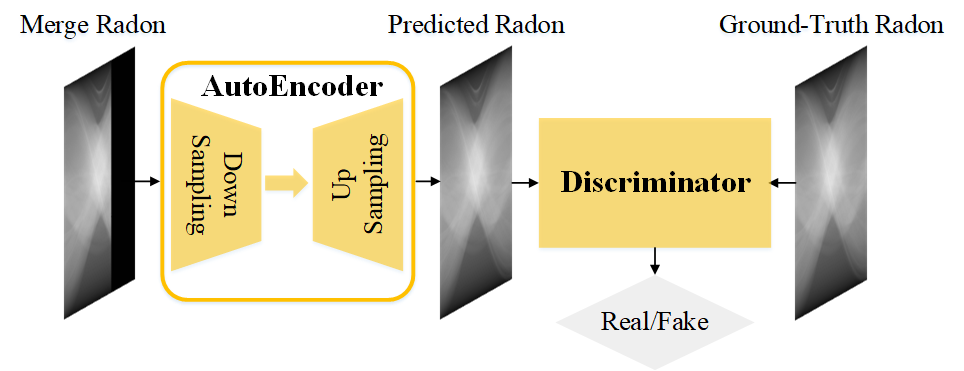}}
		\caption{The overall architecture of our proposed adversarial autoencoder in stage one.}
		\label{sec:fig3-2}
	\end{figure}

	For the input limited-view Radon data, we need to apply it as the prior information to perform angle completion in the first stage. Due to the fact that U-Net \cite{ronneberger2015u} is widely use in medical imaging currently, we propose an adversarial autoencoder with U-Net as the backbone. Its overall architecture is shown in Fig.\ref{sec:fig3-2}.

	\begin{figure}[htbp]
		
		\centerline{\includegraphics[width=\columnwidth]{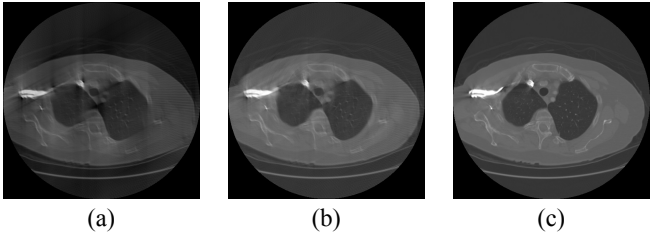}}
		\caption{(a) is the result of reconstructing the original limited-view Radon data into the image; (b) is the result of reconstructing the full-view Radon data generated from stage one into the image; (c) is the ground truth of the image.}
		\label{sec:fig3-3}
	\end{figure}
	
	We modified U-Net as the autoencoder in our adversarial autoencoder, which includes an encoder that downsamples the image to extract the representative feature and a decoder that upsamples the feature to restore the image. The precise structure of our autoencoder can be seen from TABLE I, where (Ic, Oc) represents the in-channel and out-channel of the convolutional layer. In its convolutional layers, the kernel size is 3$\times$3, the stride and padding are both 1, and the kernel size is 2$\times$2 in all of its pooling layers. In all of its deconvolution layers, the kernel size is 2$\times$2, and the stride is 1. In order to upgrade the model's ability of restoration, we combine this autencoder with a discriminator whose structure is the same as the encoder shown in TABLE I. As can be seen from Sec IV, adding this discriminator can effectively improve the model’s performance.
	
	\begin{figure*}[bp]
		\vspace{-5pt}
		\centering
		\includegraphics[scale=0.35]{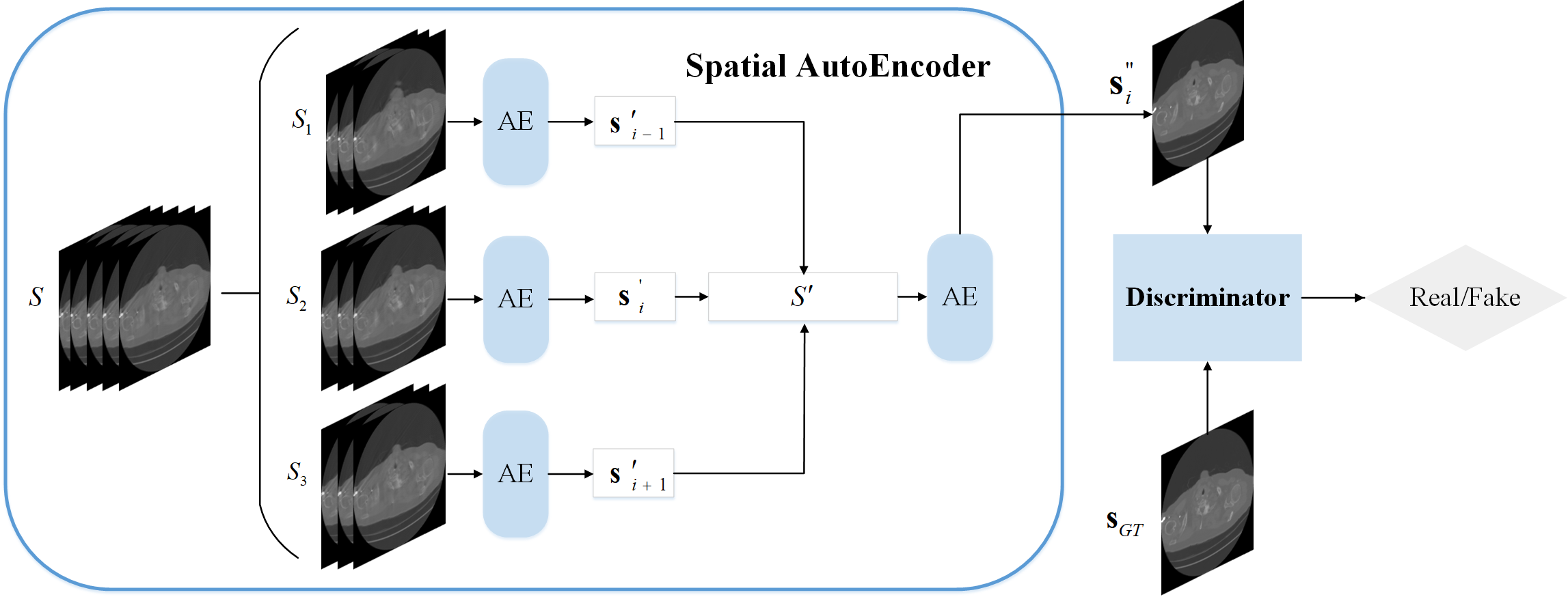}
		\caption{The overall architecture of our proposed Spatial-AAE model in stage two.}
		\label{sec:fig3-4}
		\vspace{-15pt}
	\end{figure*}
	
	\subsubsection{Stage 2: Image Restoration Based on Spatial Information}
	Fig.\ref{sec:fig3-3} shows that after we reconstruct the output from stage one into the image domain, the image texture can be partly restored, while there are still some artifacts and blurry area that can bring severe obstacles for clinical diagnosis. Therefore, in the second stage, we propose the Spatial-AAE model based on the spatial correlation between consecutive image slices to significantly improve the quality of damaged image.

	According to our knowledge, in previous studies of CT imaging restoration and reconstruction algorithms, scholars seemed to neglect the rich spatial information between consecutive image slices, and only repaired and reconstructed images in two-dimensional space. During the process of investigating and comparing the fields of image inpainting and video inpainting, we were surprised to find that the third dimension usually contains rich data coherence and continuity, which is very beneficial for restoring successive images. Thus, we suppose that the effective use of the third-dimensional information may remarkably improve the quality of restored images. Inspired by the utilization of the third-dimensional information in FastDVDNet [46], we come up with the Spatial-AAE network, whose overall architecture is shown in Fig.\ref{sec:fig3-4}, it can be divided into Spatial autoencoder and discriminator.
	
	\begin{table}[tbp]
		\vspace{-10pt}
		\caption{Details of the Autoencoder in Stage One}
		\centering
		\label{sec:tab1}
		\begin{tabular}{ccrcc}
			\cmidrule[0.75pt]{1-2}\cmidrule[0.75pt]{4-5}\morecmidrules\cmidrule[0.75pt]{1-2}\cmidrule[0.75pt]{4-5}
			\multicolumn{2}{c}{(a) Encoder} & \multirow{16}[4]{*}{} & \multicolumn{2}{c}{(b) Decoder} \\ 
			\cmidrule{1-2}\cmidrule{4-5}  \multicolumn{1}{c}{Layer}& \multicolumn{1}{c}{(Ic, Oc)} &       & \multicolumn{1}{c}{Layer} & \multicolumn{1}{c}{(Ic, Oc)} \\ \cmidrule{1-2}\cmidrule{4-5}
			\multicolumn{1}{p{4.19em}}{Conv1\_1} & \multicolumn{1}{c}{(1, 32)} &       & UpConv6 & (512, 256) \\
			\multicolumn{1}{p{4.19em}}{Conv1\_2} & \multicolumn{1}{c}{(32, 32)} &       & Concat & [UpConv6, Conv4] \\
			\multicolumn{1}{c}{Pool1} & \multicolumn{1}{c}{Maxpool} &       & Conv6\_1 & (512, 256) \\
			\multicolumn{1}{p{4.19em}}{Conv2\_1} & \multicolumn{1}{c}{(32, 64)} &       & Conv6\_2 & (256, 256) \\
			\multicolumn{1}{p{4.19em}}{Conv2\_2} & \multicolumn{1}{c}{(64, 64)} &       & UpConv7 & (256, 128) \\
			\multicolumn{1}{c}{Pool2} & \multicolumn{1}{c}{Maxpool} &       & Concat & [UpConv7, Conv3] \\
			\multicolumn{1}{p{4.19em}}{Conv3\_1} & \multicolumn{1}{c}{(64, 128)} &       & Conv7\_1 & (256, 128) \\
			\multicolumn{1}{p{4.19em}}{Conv3\_2} & \multicolumn{1}{c}{(128, 128)} &       & Conv7\_2 & (128, 128) \\
			\multicolumn{1}{c}{Pool3} & \multicolumn{1}{c}{Maxpool} &       & UpConv8 & (128, 64) \\
			\multicolumn{1}{p{4.19em}}{Conv4\_1} & \multicolumn{1}{c}{(128, 256)} &       & Concat & [UpConv8, Conv2] \\
			\multicolumn{1}{p{4.19em}}{Conv4\_2} & \multicolumn{1}{c}{(256, 256)} &       & Conv8\_1 & (128, 64) \\
			\multicolumn{1}{c}{Pool4} & \multicolumn{1}{c}{Maxpool} &       & Conv8\_2 & (64, 64) \\
			\multicolumn{1}{p{4.19em}}{Conv5\_1} & \multicolumn{1}{c}{(256, 512)} &       & UpConv9 & (64, 32) \\
			\multicolumn{1}{p{4.19em}}{Conv5\_2} & \multicolumn{1}{c}{(512, 512)} &       & Concat & [UpConv9, Conv1] \\
			&       &       & Conv9\_1 & (64, 32) \\
			&       &       & Conv9\_2 & (32, 12) \\
			&       &       & Conv9\_3 & (12, 1) \\
			\cmidrule[0.75pt]{1-2}\cmidrule[0.75pt]{4-5}\morecmidrules\cmidrule[0.75pt]{1-2}\cmidrule[0.75pt]{4-5}    \end{tabular}%
	\end{table}
	
	The input of the spatial autoencoder is five consecutive image slices $S=\{\boldsymbol{s}_{i-2},\boldsymbol{s}_{i-1},\boldsymbol{s}_i,\boldsymbol{s}_{i+1},\boldsymbol{s}_{i+2}\}$, we divide them into three sets of data $S_1=\{\boldsymbol{s}_{i-2},\boldsymbol{s}_{i-1},\boldsymbol{s}_i\},S_2=\{\boldsymbol{s}_{i-1},\boldsymbol{s}_i,\boldsymbol{s}_{i+1}\}$ and $S_3=\{\boldsymbol{s}_{i},\boldsymbol{s}_{i+1},\boldsymbol{s}_{i+2}\}$. Then, they are sent into the AE block respectively, and their output is concatenated as $S'=\{\boldsymbol{s}'_{i-1},\boldsymbol{s}'_{i},\boldsymbol{s}'_{i+1}\}$, this set of data is input into the AE block again to obtain the final restored result. The spatial autoencoder network can be expressed as (1), where $F$ is the spatial autoencoder model and $G$ is the AE block. The specific details of the AE block and discriminator in Fig.\ref{sec:fig3-4} can be seen from TABLE \ref{sec:tab1}, they are the same as they are in the AAE model of stage one.
	\begin{equation}
	\boldsymbol{s}''_i = F(S)=G\left(G(S_1),G(S_2),G(S_3)\right)
	\label{sec:equ1}
	\end{equation}
	
	\subsubsection{Stage 3: Image Refining on Patches}
\begin{figure}[htbp]
		\centerline{\includegraphics[width=\columnwidth]{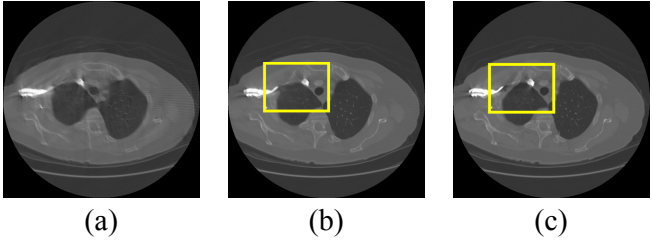}}
		\caption{(a) contains the result of reconstructing the full-view Radon data from stage one output into images; (b) contains the result of reconstructing the full-view Radon data from stage two output into images; (c) contains the ground truth of images.}
		\label{sec:fig3-5}
	\end{figure}

	It can be seen from Fig.\ref{sec:fig3-5} that after the above two stages of dual-domain combined inpainting and reconstruction, the original limited-view Radon data can be restored to a relatively satisfying extent. However, the overall details are still not precise enough.
	
\begin{figure*}[bp]
		\centering
		\includegraphics[scale=0.5]{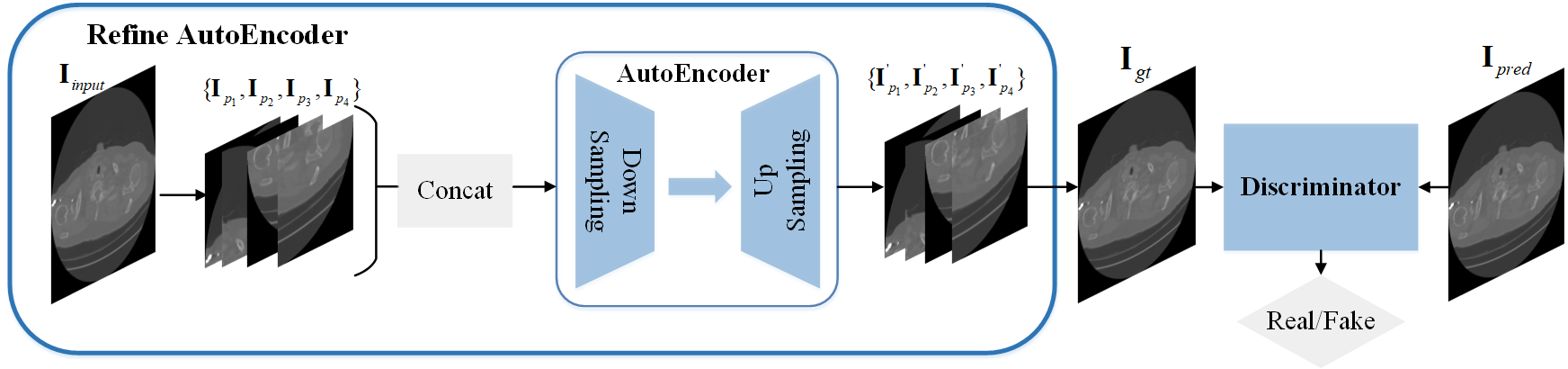}
		\caption{The overall architecture of our proposed Refine-AAE model in stage three.}
		\label{sec:fig3-6}
		\vspace{-10pt}
	\end{figure*}

	Therefore, in the third stage, we utilize the idea of “coarse to fine” in deep learning to propose the Refine-AAE model, so as to further refine the texture of repaired images. The overall structure of the Refine-AAE network can be seen from Fig.\ref{sec:fig3-6}. Give the input image $\boldsymbol{I}_{input}$, the model divides it into four patches and concatenate them into a set of sequence $\{\boldsymbol{I}_{p1},\boldsymbol{I}_{p2},\boldsymbol{I}_{p3},\boldsymbol{I}_{p4},\}$, We send it into the autoencoder for inpainting in patches and obtain the output as $\{\boldsymbol{I}^{'}_{p1},\boldsymbol{I}^{'}_{p2},\boldsymbol{I}^{'}_{p3},\boldsymbol{I}^{'}_{p4},\}$. The model integrates this output into $\boldsymbol{I}_{pred}$ and combines it with the ground truth $\boldsymbol{I}_{GT}$  into pair for discriminator’s judgment.
	
	The autoencoder and discriminator in the Refine-AAE model are the same as the Spatial-AAE model, they can be seen from TABLE I.
	
	\subsection{Loss Function}
	In all three stages, we use multi-loss function to optimize the autoencoder model, it can be expressed as \eqref{sec:equ2}.
	
	\begin{equation}
	l_{AE}=\alpha_1l_{MSE}+\alpha_2l_{adv}+\alpha_3l_{reg}
	\label{sec:equ2}
	\end{equation}
	$l_{MSE}$ calculates the mean square error between the restored image and the ground truth image, it is widely used in various image inpainting tasks because it can provide an intuitive evaluation for the model’s prediction. The expression of $l_{MSE}$ can be seen from \eqref{sec:equ3}.
	\begin{equation}
	l_{MSE}=\dfrac{1}{W \times H}\sum_{x=1}^{W}\sum_{y=1}^{H}\left(\boldsymbol{I}_{x,y}^{GT}-G_{AE}(\boldsymbol{I}^{input})_{x,y}\right)^2
	\label{sec:equ3}
	\end{equation}
	where $G_{AE}$ is the auto-encoder, $\boldsymbol{I}^{GT}$ and $\boldsymbol{I}^{input}$ are the ground truth image and the input image, $W$ and $H$ are the width and height of the input image respectively. 
	$l_{adv}$ refers to the adversarial loss. The autoencoder can fool the discriminator by making its prediction as close to the ground truth as possible, so as to achieve the ideal image restoration outcome. Its expression can be seen from \eqref{sec:equ4}.
	\begin{equation}
	l_{adv}=1-D\left(G_{AE}(\boldsymbol{I}^{input})\right)
	\label{sec:equ4}
	\end{equation}
	where $D$ is the discriminator and $G_{AE}$ is the autoencoder.
	$l_{reg}$ is the regularization term of our multi-loss function. Since noises may have a huge impact on the restoration result, we add a regularization term to maintain the smoothness of the image and also avoid the problem of overfitting. TV Loss is commonly used in image analysis tasks, it can reduce the difference between adjacent pixel values in the image to a certain extent. Its expression can be seen from \eqref{sec:equ5}.
	\begin{equation}
	l_{reg}=\dfrac{1}{W \times H}\sum_{x=1}^{W}\sum_{y=1}^{H}\left\|\nabla G_{AE}(\boldsymbol{I}^{input}_{x,y}) \right\|
	\label{sec:equ5}
	\end{equation}
	where $G_{AE}$ is the auto-encoder, $\boldsymbol{I}_{input}$ is the input image, $W$ and $H$ are the width and height of the input image respectively. $\nabla$ calculates the gradient, $\left\|   \right\|$ calculates the norm.
	
	For the optimization of the discriminator, the loss function should enable the discriminator to better distinguish between real and fake inputs. The loss function can be seen from \eqref{sec:equ6}.
	\begin{equation}
	l_{DIS}=1-D(\boldsymbol{I}^{GT})+D\left(G_{AE}(\boldsymbol{I}^{input})\right)
	\label{sec:equ6}
	\end{equation}
	where $D$ is the discriminator, $G^{AE}$ is the auto-encoder, $\boldsymbol{I}^{GT}$ and $\boldsymbol{I}^{input}$ are the ground truth image and the input image respectively. The discriminator outputs a scalar between zero and one, when the output is closer to 1, the discriminator thinks that the input is more likely to be real. On the opposite, when the output is closer to 0, it thinks the input is more likely to be fake. Therefore, $1-D(\boldsymbol{I}^{GT})$ makes the output closer to one when the discriminator inputs real images, and $D\left(G_{AE}(\boldsymbol{I}^{input})\right)$  makes the output closer to zero when the discriminator inputs fake images generated by the autoencoder.
	
	\section{Experiment}
	Our experiment data comes from 1000 cases in the LIDC-IDRI \cite{armato2010we} dataset. We divided cases 1 to 200 into the test set, cases 201 to 400 into the validation set, and cases 401 to 1000 into the training set. The CT imaging (size 512$\times$512) is stored as DCM files in the LIDC-IDRI dataset. After processing it as an array, we reconstruct it to the Radon domain (size 512$\times$180) through the Radon transformation, and perform post-60-degree clipping on it as the input data of the overall model. During the training process, we set the learning rate to 1e-4, using ADAM \cite{kingma2014adam} as our model optimizer, and Leaky ReLU \cite{maas2013rectifier} as the nonlinear activation. For the multi-loss function, we refer to the method in paper \cite{ledig2017photo}, where $\alpha_1$, $\alpha_2$ and $\alpha_3$ are set to 1, 1e-3, and 2e-8 respectively.
	It is worth mentioning that there is no fully connected layer in our model, so it can flexibly handle input images of different resolutions and be applied to various datasets. In addition, unlike other deep learning-based algorithms, the reconstruction part of our algorithm adopts FBP instead of SART-TV which requires a relatively high level of computational complexity, so our method can be better applied to practical application scenarios such as clinical diagnosis. Although FBP takes much shorter time than SART-TV, its reconstruction results have a certain gap with SART-TV. In order to realize the practical application value of our algorithm, we manage to compensate the performance of FBP through the superiority of our model design. Also, we increase the damage degree of Radon data in 4.2 to test the robustness of our algorithm. We create four types of damaged Radon data and use this algorithm to repair and reconstruct them. The experimental results prove that our algorithm can effectively restore these data, thus owns outstanding robustness.
	In 4.1, we conduct ablation experiments on models of each stage to prove the necessity and effectiveness of our structural design. In 4.2, we compared our algorithm with other four types of algorithms, and test these algorithms on four various degrees of damaged data.
	
	\subsection{Ablation Study}
	\subsubsection{Stage1}
	We first explore the necessity of fusing data in the Radon domain (refers to Fig.\ref{sec:fig3-1}). We input the directly cut Radon data and the fused Radon data into the stage one model shown in Fig.\ref{sec:fig3-2} for data completion, and compare their outputs with the Radon ground truth. The experimental results can be seen in TABLE II, CR stands for the directly cut Radon, MR stands for the merged Radon, RCR stands for the restored CR from stage1, RMR stands for the restored MR from stage1.
	
\begin{figure}[htbp]
		\centerline{\includegraphics[width=\columnwidth]{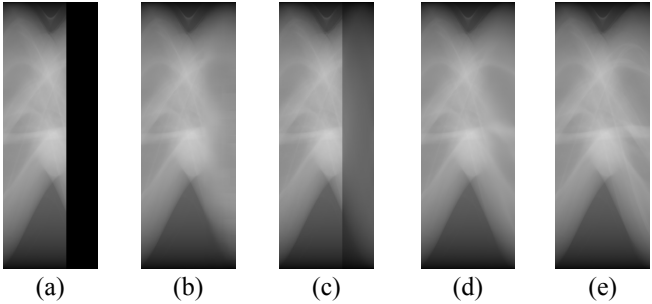}}
		\caption{Visualized results obtained from different data preprocessing methods, (a) is the directly cut Radon data; (b) is the restored result of (a); (c) is the fused Radon data; (d) is the restored result of (c); (e) is the Radon ground truth.}
		\label{sec:fig4-1}
	\end{figure}

	It can be concluded from TABLE II that the fused Radon data can obtain better experimental results due to its richer prior information, and provide more texture for the subsequent image restoration steps. The visualized results can be seen in Fig.\ref{sec:fig4-1}.
	
	\begin{table}[htbp]
		\centering
		\caption{Details of Various Data Preprocessing Methods}
		\label{sec:tab2}
		\begin{tabular}{m{30pt}m{35pt}<{\centering}m{35pt}<{\centering}m{35pt}<{\centering}m{35pt}<{\centering}}
			\toprule\toprule
			& CR & MR & RCR & \textbf{RMR} \\ \midrule
			PSNR & 8.714     & 18.196      & 38.549             & \textbf{48.181}               \\ 
			SSIM & 0.656     & 0.936       & 0.987              & \textbf{0.995}                \\ 
			\bottomrule 
			\bottomrule  	
		\end{tabular}	
	\end{table}

	In addition, we also explore the architecture of stage one's adversarial autoencoder model, and proved that it is essential to add the discriminator reasonably. We restore the input data with: (1) The autoencoder shown in TABLE I (a); (2) Combination of the autoencoder and the discriminator in TABLE I, their experimental results can be seen from TABLE III.
	
	\begin{table}[htbp]
		\centering
		\caption{Results of Using Diffrent Model Structure in Stage One}
		\label{sec:tab3}
		\begin{tabular}{m{35pt}<{\centering}m{35pt}<{\centering}m{35pt}<{\centering}}
			\toprule\toprule
			& AE     & \textbf{AE + D} \\ \midrule
			PSNR & 40.129 &\textbf{48.181} \\ 
			SSIM & 0.983  & \textbf{0.995}  \\ 
			\bottomrule 
			\bottomrule 
		\end{tabular}
	\end{table}
	
	It can be summarized from the above data that adding a discriminator can greatly improve the data completion result. It can help stage one model to improve the sinogram data PSNR by a relatively large margin.

	\begin{figure}[htbp]
		
		\centerline{\includegraphics[scale=0.35]{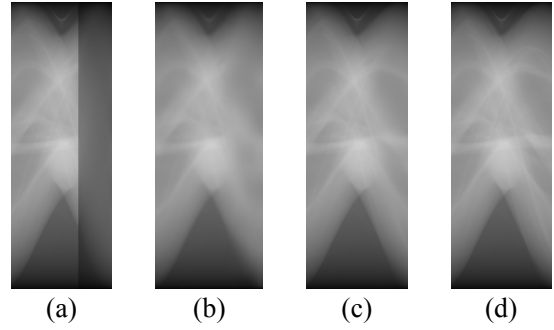}}
		\caption{Visualized results obtained from different model structure in stage one, (a) is the input fused Radon data; (b) is the restored Radon data from structure AE; (c) is the restored Radon data from structure AE+D; (d) is the Radon ground truth.}
		\label{sec:fig4-2}
	\end{figure}
	
	From the visualized comparison in Fig.\ref{sec:fig4-2}, we can see that if we only use this single autoencoder, the inpainting result would have a large blurred area, and adding the discriminator can improve this situation.
	
	\subsubsection{STAGE2}
	For the image restoration task in this stage, we adopt the Spatial-AAE model described in 3.2 to make full use of spatial information. In order to reflect the prominence of this structure, we compare this model with the AAE model from stage one, which does not contain any spatial structure. For the same input fused Radon data, the experimental results can be seen in TABLE IV.
	It can be seen from the results that, due to the fact that the Spatial-AAE model makes full use of the third-dimensional prior information, it can effectively improve the overall performance of stage two.
	
	\begin{table}[htbp]
		\centering
		\caption{Results of Using Different Model Structure in Stage Two}
		\label{sec:tab4}
		\begin{tabular}{m{30pt}<{\centering}m{35pt}<{\centering}m{50pt}<{\centering}}
			\toprule\toprule
			& AAE & \textbf{Spacial-AAE} \\ \midrule
			PSNR & 37.384 & \textbf{39.646} \\
			SSIM & 0.929 & \textbf{0.940} \\ 
			\bottomrule 
			\bottomrule 
		\end{tabular}
	\end{table}	
	
	\subsubsection{STAGE3}
\begin{figure}[htbp]
		\centerline{\includegraphics[width=\columnwidth]{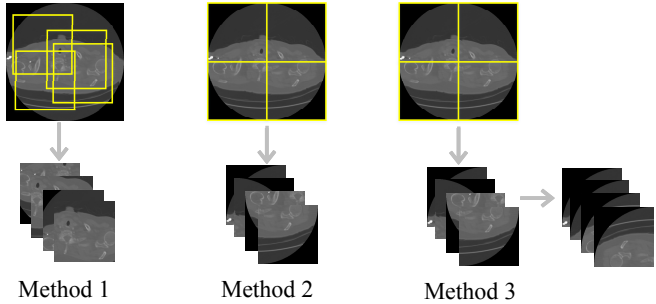}}
		\caption{Methods of cropping patches in stage three}
		\label{sec:fig4-3}
	\end{figure}

	In this stage, the input image is divided and concatenated, and then sent to the Refine-AAE model for finer inpainting. We believe that the way of intercepting patches during the training process will have a certain impact on the experimental results, so we test the following three types of interception methods (As shown in Fig.\ref{sec:fig4-3}): (1) Randomly crop four patches (size 256$\times$256) from the input image (size 512$\times$512); (2) Crop the four corners out of the input image; (3) Crop the four corners out of the input image, and then adjust them into the same pattern through different flipping method.
	
	All of the methods above get an array of size (4, 256, 256), we input it into the Refine-AAE model (refers to Fig.\ref{sec:fig3-5}) to finely repair the image, and the experimental results of these three methods are shown in TABLE V.
	
	\begin{table}[htbp]
		\centering
		\caption{Results of Using Patch-Cropping Methods in Stage Three}
		\label{sec:tab5}
		\begin{tabular}{cccc}
			\toprule\toprule
			& Random Crop & \textbf{Corner Crop} & Corner Crop + Flip \\ \midrule
			PSNR & 40.111      & \textbf{40.209}      & 40.060             \\ 
			SSIM & 0.942       & \textbf{0.943}       & 0.942              \\ \bottomrule 
			\bottomrule 
		\end{tabular}
	\end{table}
	
	We can conclude that method (2) achieves the best image restoration result, this is different from our initial assumption. We originally assumed that patches generated from method (3) can enable the model to learn the mapping easier. However, the fact is that method (2) gets the better result. We suppose this is because different patterns in method (2) play a crucial role in data enhancement, thus prevent the model from overfitting.
	
	\subsection{Algorithm Comparison}
\begin{figure}[!t]
		\centering
		\includegraphics[scale=0.6]{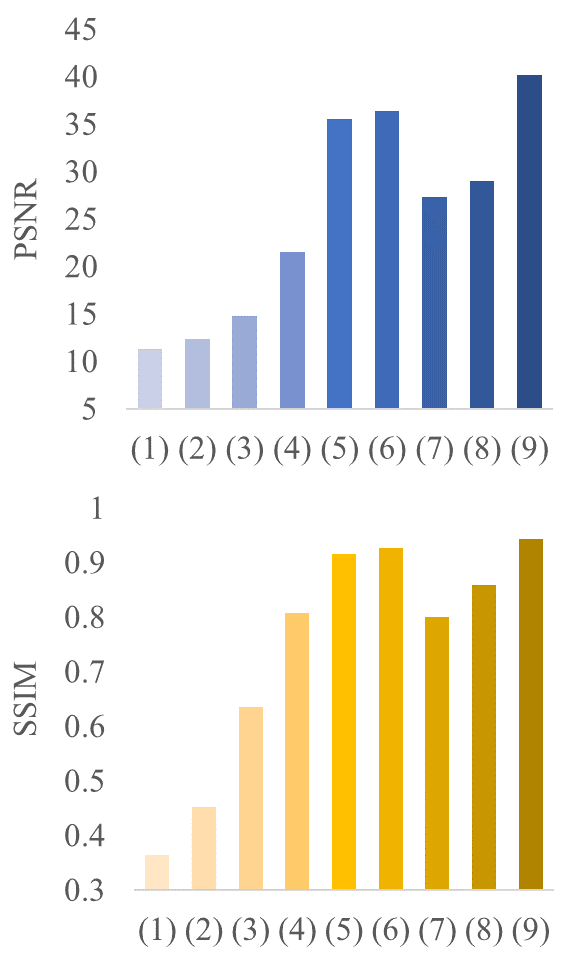}
		\vspace{-7pt}
		\caption{Histograms of different algorithms applied to different data preprocessing methods.}
		\label{sec:fig4-4}
		\vspace{-5pt}
	\end{figure}

	In order to reflect the superiority of our algorithm, we have compared its performance with the following four sorts of algorithms: (1) Analytical reconstruction algorithm FBP; (2) Iterative reconstruction algorithm SART combined with TV regularization; (3) Image inpainting, after reconstructing the limited-view Radon data into images through FBP, apply the AAE model to image restoration; (4) Sinogram inpainting, first use the AAE model to complement the Radon data, and then adopt FBP to reconstruct it to images. We also test these algorithms on two types of input data: (1) the directly cut Radon data; (2) the fused Radon data. For Radon data with its post 60 degrees being cut off, the performance of the above algorithms is shown in TABLE VI and Fig.\ref{sec:fig4-4}, MR in this table means input the fused Radon data.
	
	\begin{table}[htbp]
		\caption{Results of Different Algorithms Applied to Different Data Preprocessing Methods}
		\label{sec:tab6}
		\begin{center}
			\begin{tabular}{lm{35pt}<{\centering}m{35pt}<{\centering}}
				\toprule\toprule
				Algorithms & PSNR & SSIM \\ \midrule
				(1) FBP & 11.272 & 0.364 \\ 
				(2) FBP+MR & 12.354 & 0.452 \\ 
				(3) SART-TV & 14.727 & 0.635	 \\
				(4) SART-TV+MR & 21.518 & 0.807 \\
				(5) Image Inpainting (II) & 35.566 & 0.916 \\ 
				(6) Image Inpainting + MR & 36.388  & 0.927 \\ 
				(7) Sinogram Inpainting (SI) & 27.345 & 0.800 \\ 
				(8) Sinogram Inpainting + MR & 28.960 & 0.859 \\ 
				(9) \textbf{Ours} & \textbf{40.209} & \textbf{0.943} \\ 
				\bottomrule 
				\bottomrule 
			\end{tabular}
		\end{center}
		
	\end{table}

	\begin{figure*}[htbp]
		\centering
		\includegraphics[scale=0.3]{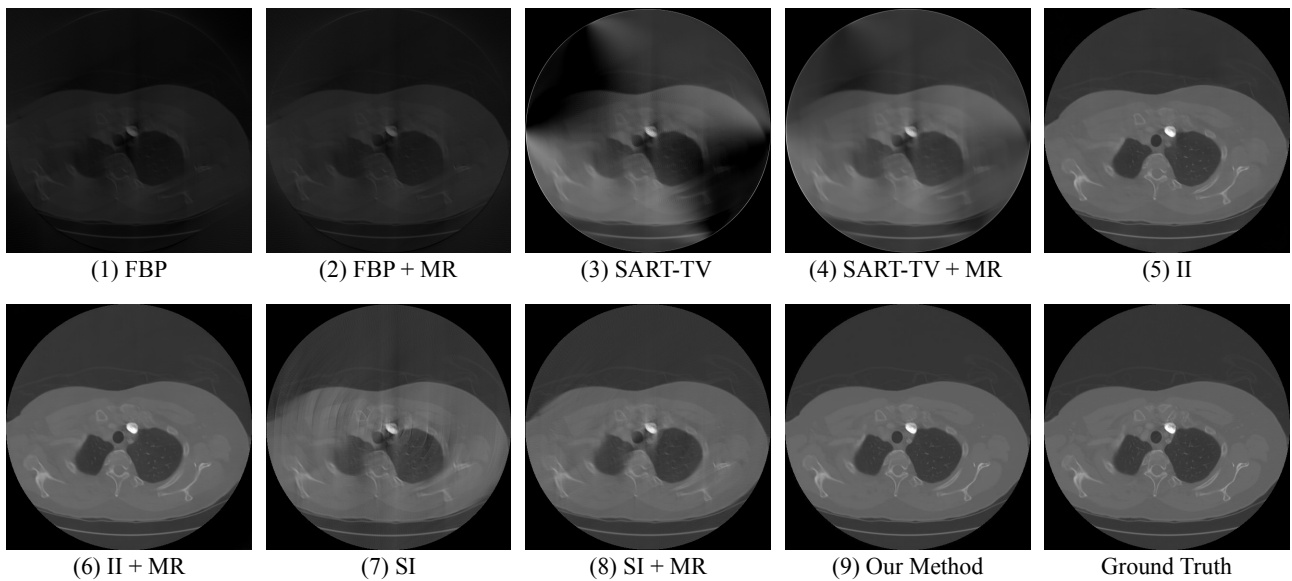}
		\vspace{-7pt}
		\caption{Visualized results of different algorithms applied to different data preprocessing methods. }
		\label{sec:fig4-5}
		\vspace{-2pt}
	\end{figure*}
	
	\begin{figure*}[hbp]
		\centering
		\includegraphics[scale=0.3]{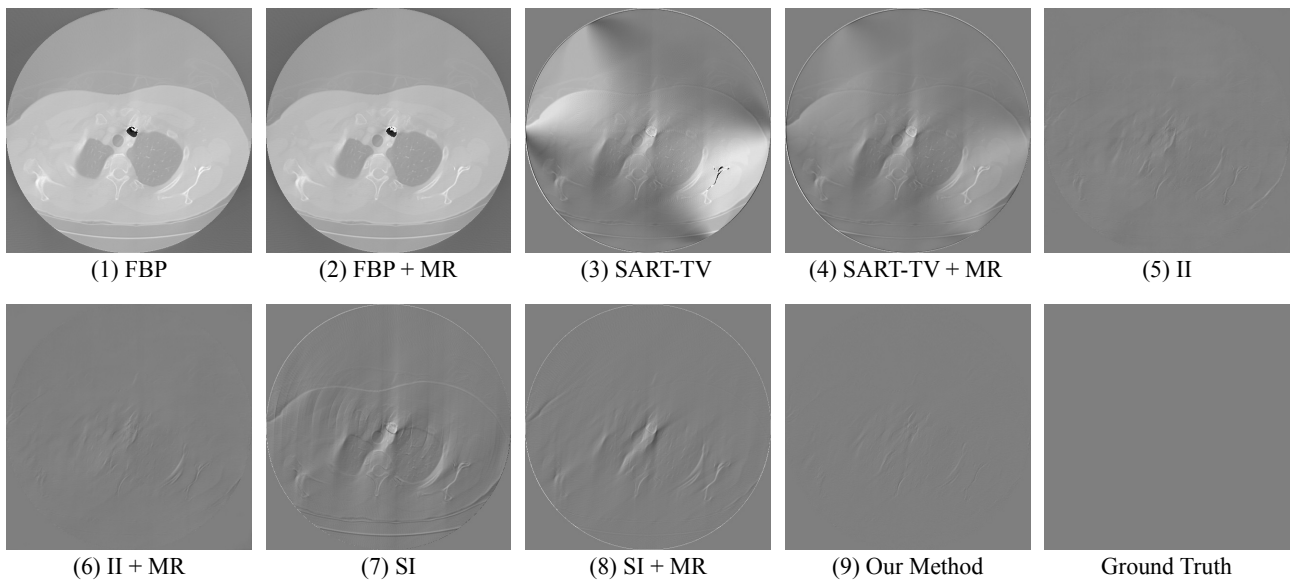}
		\vspace{-7pt}
		\caption{Error maps of different algorithms applied to different data preprocessing methods.}
		\label{sec:fig4-6}
		\vspace{-2pt}
	\end{figure*}
	
	\begin{figure*}
		\centering
		\includegraphics[scale=0.75]{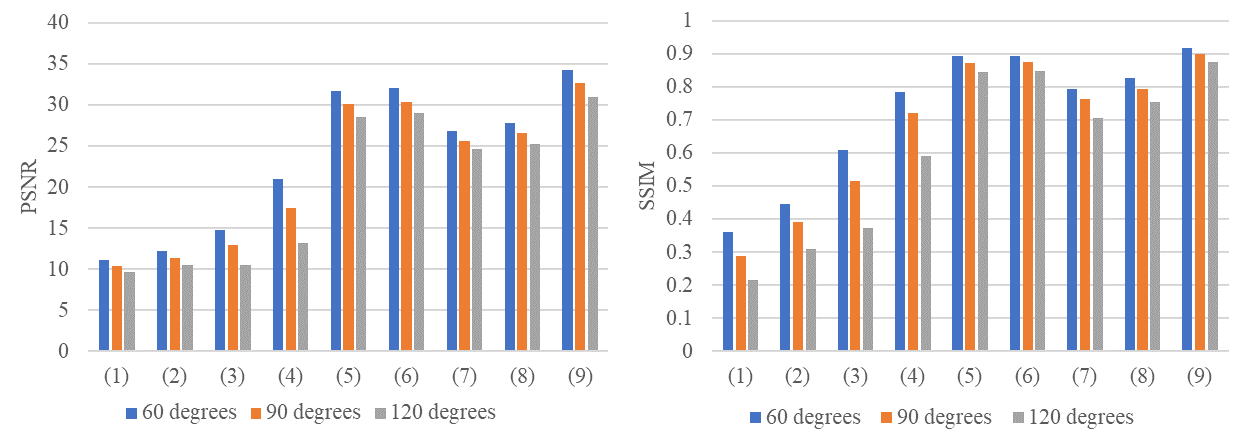}
		\caption{Histograms of different algorithms applied to different data preprocessing methods on different input data}
		\label{sec:fig4-7}
		
	\end{figure*}

	\begin{table*}[htbp]
		\centering
		
		\label{sec:tab7}
		\vspace{-2pt}
		\setlength{\tabcolsep}{6mm}{
			\begin{tabular}{lcccccc}
				\toprule\toprule
				& \multicolumn{2}{c}{CUT-MID-60} & \multicolumn{2}{c}{CUT-MID-90} & \multicolumn{2}{c}{CUT-MID-120} \\ \midrule
				Algorithms                   & PSNR            & SSIM          & PSNR            & SSIM          & PSNR            & SSIM           \\ \midrule
				(1) FBP                      & 11.131          & 0.362         & 10.350          & 0.289         & 9.636           & 0.217          \\ 
				(2) FBP + MR                 & 12.182          & 0.446         & 11.432          & 0.391         & 10.525          & 0.309          \\ 
				(3) SART-TV                  & 14.758          & 0.610         & 12.945          & 0.515         & 10.492          & 0.372          \\ 
				(4) SART-TV + MR             & 21.036          & 0.784         & 17.523          & 0.722         & 13.166          & 0.592          \\ 
				(5) Image Inpainting (II)         & 31.717          & 0.895         & 30.157          & 0.873         & 28.507          & 0.846          \\ 
				(6) Image Inpainting + MR    & 32.031          & 0.895         & 30.422          & 0.876         & 28.999          & 0.849          \\ 
				(7) Sinogram Inpainting (SI)      & 26.834          & 0.793         & 25.673          & 0.763         & 24.606          & 0.705          \\ 
				(8) Sinogram Inpainting + MR & 27.789          & 0.828         & 26.582          & 0.795         & 25.210          & 0.755          \\ 
				(9) \textbf{Ours}                     & \textbf{34.248}          & \textbf{0.919}         & \textbf{32.624}          & \textbf{0.900}         & \textbf{30.975}          & \textbf{0.876}          \\ 
				\bottomrule 
				\bottomrule 
			\end{tabular}
		\vspace{-2pt}
		\caption{Results of Different Algorithms Applied to Different Data Preprocessing Methods on Different Input Data}
		\vspace{-5pt}
		}
		
	\end{table*}
	
	From the results above, we can see that the idea of merging Radon data brings additional prior information on every type of algorithm, thus improves their performance by different margin. Besides, merging Radon data is particularly helpful for SART-TV. Under the condition of using the same AAE model, restoration in the image domain is more effective than restoration in the Radon domain. Our algorithm combines the Radon domain and the image domain, complements, reconstructs, restores and refines the input limited-view Radon data, can finally improves the image PSNR to 40.209 and SSIM to 0.943. It upgrades the quality of CT imaging by a large margin, realizes the accurate restoration of its texture. Comparison of the visualized results can be seen in Fig.\ref{sec:fig4-5}, we also present the corresponding error maps in Fig.\ref{sec:fig4-6} to reflect the difference in performance between different algorithms.
	To prove the robustness of our algorithm, we test various degrees of damage on the input Radon data, including (1) cut off the middle 60 degrees (1/3 of the original data); (2) cut off the middle 90 degrees (1/2 of the original data); (3) cut off the middle 120 degrees (2/3 of the original data). We implement the above nine algorithms in TABLE VI on these three types of input data, and their performance is shown in TABLE VII and Fig.\ref{sec:fig4-7}.
	It can be seen that losing data in the middle can cause more damage than in the rear. With the increase of the cropping ratio, the inpainting performance of these algorithms has also been greatly affected. Our algorithm however, proves its outstanding robustness under various conditions. Even when cutting the middle 120 degrees off the original Radon data, our method can still restore the seriously damaged imaging to PSNR of 30.975. Also, our method can exceed the other methods in TABLE VII by a large margin under varying degrees of damaged data.
	
	\section{Conclusion}
	In order to improve the quality of the seriously damaged limited-view CT imaging, we propose a three-stage restoration and reconstruction algorithm based on spatial information, which combines the Radon domain and the image domain, and utilizes the idea of “coarse to fine” to restore the image with high definition. In the first stage, we designed an adversarial autoencoder to complement the limited-view Radon data. In the second stage, we first reconstruct the Radon data into images through FBP, and then send this image into the Spatial-AAE model we built to achieve image artifact and noise reduction based on spatial correlation between consecutive slices. In the third stage, we propose the Refine-AAE network to finely repair the image in patches, so as to achieve the accurate restoration of the image texture. For Radon data with limited angle of 120 degrees (cut off one-third of the full-view Radon data), our algorithm can increase its PSNR to 40.209, and SSIM to 0.943. At the same time, due to the fact that our model does not restrict input resolution, can adapt to varying degrees of damage, and also can be quickly implemented, our algorithm has generalization, robustness and significant practical application value.
	
	In our future work, we hope to incorporate our three-stage model into an end-to-end network that can be simultaneously trained and tested. As we all know, such large amount of parameters may be hard to optimize, we plan to solve this problem by using tricks such as data augmentation and dropout, while also lightweight model backbone like MobileNet.
	
	\bibliography{IEEEabrv,Ref}

\begin{thebibliography}{10}
\providecommand{\url}[1]{#1}
\csname url@samestyle\endcsname
\providecommand{\newblock}{\relax}
\providecommand{\bibinfo}[2]{#2}
\providecommand{\BIBentrySTDinterwordspacing}{\spaceskip=0pt\relax}
\providecommand{\BIBentryALTinterwordstretchfactor}{4}
\providecommand{\BIBentryALTinterwordspacing}{\spaceskip=\fontdimen2\font plus
\BIBentryALTinterwordstretchfactor\fontdimen3\font minus
  \fontdimen4\font\relax}
\providecommand{\BIBforeignlanguage}[2]{{%
\expandafter\ifx\csname l@#1\endcsname\relax
\typeout{** WARNING: IEEEtran.bst: No hyphenation pattern has been}%
\typeout{** loaded for the language `#1'. Using the pattern for}%
\typeout{** the default language instead.}%
\else
\language=\csname l@#1\endcsname
\fi
#2}}
\providecommand{\BIBdecl}{\relax}
\BIBdecl

\bibitem{chen2017low}
H.~Chen \emph{et~al.}, ``{Low-dose CT with a residual encoder-decoder
  convolutional neural network},'' \emph{IEEE Trans. Med. Imaging}, vol.~36,
  no.~12, pp. 2524--2535, 2017.

\bibitem{kalra2004strategies}
M.~K. Kalra \emph{et~al.}, ``{Strategies for CT radiation dose optimization},''
  \emph{Eur. J. Radiol.}, vol. 230, no.~3, pp. 619--628, 2004.

\bibitem{slovis2002alara}
T.~L. Slovis, ``{The ALARA concept in pediatric CT: myth or reality?}''
  \emph{Eur. J. Radiol.}, vol. 223, no.~1, pp. 5--6, 2002.

\bibitem{khare2005a}
A.~Khare and U.~S. Tiwary, ``A new method for deblurring and denoising of
  medical images using complex wavelet transform,'' in \emph{2005 IEEE
  Engineering in Medicine and Biology 27th Annual Conference}, vol.~2, 2005,
  pp. 1897--1900.

\bibitem{xie2018artifact}
S.~Xie \emph{et~al.}, ``{Artifact removal using improved GoogLeNet for
  sparse-view CT reconstruction},'' \emph{Sci. Rep.}, vol.~8, no.~1, pp. 1--9,
  2018.

\bibitem{zhang2019dualres}
T.~Zhang, H.~Gao, Y.~Xing, Z.~Chen, and L.~Zhang, ``{DualRes-UNet: Limited}
  angle artifact reduction for computed tomography,'' in \emph{2019 IEEE
  Nuclear Science Symposium and Medical Imaging Conference (NSS/MIC)}.\hskip
  1em plus 0.5em minus 0.4em\relax IEEE, 2019, pp. 1--3.

\bibitem{xie2019artifact}
S.~Xie, H.~Xu, and H.~Li, ``Artifact removal using gan network for
  limited-angle {CT} reconstruction,'' in \emph{2019 Ninth International
  Conference on Image Processing Theory, Tools and Applications (IPTA)}.\hskip
  1em plus 0.5em minus 0.4em\relax IEEE, 2019, pp. 1--4.

\bibitem{zhang2020artifact}
Q.~Zhang, Z.~Hu, C.~Jiang, H.~Zheng, Y.~Ge, and D.~Liang, ``Artifact removal
  using a hybrid-domain convolutional neural network for limited-angle computed
  tomography imaging,'' \emph{Phys. Med. Biol.}, 2020.

\bibitem{xie2012image}
J.~Xie, L.~Xu, and E.~Chen, ``Image denoising and inpainting with deep neural
  networks,'' in \emph{Advances in neural information processing systems},
  2012, pp. 341--349.

\bibitem{katsevich2002theoretically}
A.~Katsevich, ``{Theoretically exact filtered backprojection-type inversion
  algorithm for spiral CT},'' \emph{SIAM J. Appl. Math.}, vol.~62, no.~6, pp.
  2012--2026, 2002.

\bibitem{imai2009statistical}
K.~Imai, M.~Ikeda, Y.~Enchi, and T.~Niimi, ``{Statistical characteristics of
  streak artifacts on CT images: Relationship between streak artifacts and mA s
  values},'' \emph{Med. Phys.}, vol.~36, no.~2, pp. 492--499, 2009.

\bibitem{liu2014model}
L.~Liu, ``Model-based iterative reconstruction: a promising algorithm for
  today's computed tomography imaging,'' \emph{Journal of Medical imaging and
  Radiation sciences}, vol.~45, no.~2, pp. 131--136, 2014.

\bibitem{chen2008prior}
G.~H. Chen, J.~Tang, and S.~Leng, ``{Prior image constrained compressed sensing
  (PICCS): a method to accurately reconstruct dynamic CT images from highly
  undersampled projection data sets},'' \emph{Med. Phys.}, vol.~35, no.~2, pp.
  660--663, 2008.

\bibitem{chen2011time}
G.~H. Chen \emph{et~al.}, ``{Time-resolved interventional cardiac C-arm
  cone-beam CT: An application of the PICCS algorithm},'' \emph{IEEE Trans.
  Med. Imaging}, vol.~31, no.~4, pp. 907--923, 2011.

\bibitem{rudin1992nonlinear}
L.~I. Rudin, S.~Osher, and E.~Fatemi, ``Nonlinear total variation based noise
  removal algorithms,'' \emph{Physica D}, vol.~60, no. 1-4, pp. 259--268, 1992.

\bibitem{sidky2008image}
E.~Y. Sidky and X.~Pan, ``Image reconstruction in circular cone-beam computed
  tomography by constrained, total-variation minimization,'' \emph{Phys. Med.
  Biol.}, vol.~53, no.~17, p. 4777, 2008.

\bibitem{niu2014sparse}
S.~Niu \emph{et~al.}, ``{Sparse-view x-ray CT reconstruction via total
  generalized variation regularization},'' \emph{Phys. Med. Biol.}, vol.~59,
  no.~12, p. 2997, 2014.

\bibitem{sidky2006accurate}
E.~Y. Sidky, C.-M. Kao, and X.~Pan, ``{Accurate image reconstruction from
  few-views and limited-angle data in divergent-beam CT},'' \emph{J. X-ray Sci.
  Technol.}, vol.~14, no.~2, pp. 119--139, 2006.

\bibitem{andersen1984simultaneous}
A.~H. Andersen and A.~C. Kak, ``{Simultaneous algebraic reconstruction
  technique (SART): a superior implementation of the ART algorithm},''
  \emph{Ultrasonic imaging}, vol.~6, no.~1, pp. 81--94, 1984.

\bibitem{xu2012low}
Q.~Xu, H.~Yu, X.~Mou, L.~Zhang, J.~Hsieh, and G.~Wang, ``{Low-dose X-ray CT
  reconstruction via dictionary learning},'' \emph{IEEE Trans. Med. Imaging},
  vol.~31, no.~9, pp. 1682--1697, 2012.

\bibitem{cao2013limited}
M.~Cao and Y.~Xing, ``Limited angle reconstruction with two dictionaries,'' in
  \emph{2013 IEEE Nuclear Science Symposium and Medical Imaging Conference
  (2013 NSS/MIC)}.\hskip 1em plus 0.5em minus 0.4em\relax IEEE, 2013, pp. 1--4.

\bibitem{zhang2017low}
H.~Zhang, L.~Zhang, Y.~Sun, J.~Zhang, and L.~Chen, ``{Low dose CT image
  statistical iterative reconstruction algorithms based on off-line dictionary
  sparse representation},'' \emph{Optik}, vol. 131, pp. 785--797, 2017.

\bibitem{xu2019l0dl}
M.~Xu, D.~Hu, and W.~Wu, ``$\ell$0dl: {Joint} image gradient $\ell$0-norm with
  dictionary learning for limited-angle {CT},'' in \emph{Proceedings of the
  10th ACM International Conference on Bioinformatics, Computational Biology
  and Health Informatics}, 2019, pp. 538--538.

\bibitem{lecun2015deep}
Y.~LeCun, Y.~Bengio, and G.~Hinton, ``Deep learning,'' \emph{nature}, vol. 521,
  no. 7553, pp. 436--444, 2015.

\bibitem{he2016deep}
K.~He, X.~Zhang, S.~Ren, and J.~Sun, ``Deep residual learning for image
  recognition,'' in \emph{Proceedings of the IEEE conference on computer vision
  and pattern recognition}, 2016, pp. 770--778.

\bibitem{srivastava2015training}
R.~K. Srivastava, K.~Greff, and J.~Schmidhuber, ``Training very deep
  networks,'' in \emph{Advances in neural information processing systems},
  2015, pp. 2377--2385.

\bibitem{dong2015image}
C.~Dong, C.~C. Loy, K.~He, and X.~Tang, ``Image super-resolution using deep
  convolutional networks,'' \emph{IEEE Trans. Pattern Anal. Mach. Intell.},
  vol.~38, no.~2, pp. 295--307, 2015.

\bibitem{wurfl2016deep}
T.~W{\"u}rfl, F.~C. Ghesu, V.~Christlein, and A.~Maier, ``Deep learning
  computed tomography,'' in \emph{International conference on medical image
  computing and computer-assisted intervention}.\hskip 1em plus 0.5em minus
  0.4em\relax Springer, 2016, pp. 432--440.

\bibitem{mao2016image}
X.~Mao, C.~Shen, and Y.-B. Yang, ``Image restoration using very deep
  convolutional encoder-decoder networks with symmetric skip connections,'' in
  \emph{Advances in neural information processing systems}, 2016, pp.
  2802--2810.

\bibitem{zhang2016image}
\BIBentryALTinterwordspacing
H.~Zhang \emph{et~al.} (2016) Image prediction for limited-angle tomography via
  deep learning with convolutional neural network. [Online]. Available:
  \url{https://arxiv.org/abs/1607.08707}
\BIBentrySTDinterwordspacing

\bibitem{kang2017deep}
E.~Kang, J.~Min, and J.~C. Ye, ``{A deep convolutional neural network using
  directional wavelets for low-dose X-ray CT reconstruction},'' \emph{Med.
  Phys.}, vol.~44, no.~10, pp. e360--e375, 2017.

\bibitem{zhang2018sparse}
Z.~Zhang, X.~Liang, X.~Dong, Y.~Xie, and G.~Cao, ``{A sparse-view CT
  reconstruction method based on combination of DenseNet and deconvolution},''
  \emph{IEEE Trans. Med. Imaging}.

\bibitem{wang2020deep}
J.~Wang, J.~Liang, J.~Cheng, Y.~Guo, and L.~Zeng, ``Deep learning based image
  reconstruction algorithm for limited-angle translational computed
  tomography,'' \emph{PLoS One}, vol.~15, no.~1, p. e0226963, 2020.

\bibitem{ronneberger2015u}
O.~Ronneberger, P.~Fischer, and T.~Brox, ``{U-net: Convolutional networks for
  biomedical image segmentation},'' in \emph{International Conference on
  Medical image computing and computer-assisted intervention}.\hskip 1em plus
  0.5em minus 0.4em\relax Springer, 2015, pp. 234--241.

\bibitem{goodfellow2014generative}
Goodfellow \emph{et~al.}, ``Generative adversarial nets,'' in \emph{Advances in
  neural information processing systems}, 2014, pp. 2672--2680.

\bibitem{anirudh2019improving}
\BIBentryALTinterwordspacing
R.~Anirudh, H.~Kim, J.~J. Thiagarajan, A.~K. Mohan, and K.~M. Champley. (2019)
  {Improving Limited Angle CT Reconstruction with a Robust GAN Prior}.
  [Online]. Available: \url{https://arxiv.org/abs/1910.01634}
\BIBentrySTDinterwordspacing

\bibitem{2019A}
Z.~Li \emph{et~al.}, ``A sinogram inpainting method based on generative
  adversarial network for limited-angle computed tomography,'' in \emph{15th
  International Meeting on Fully Three-Dimensional Image Reconstruction in
  Radiology and Nuclear Medicine}, vol. 11072.\hskip 1em plus 0.5em minus
  0.4em\relax International Society for Optics and Photonics, 2019, p. 1107220.

\bibitem{Li2019Promising}
Z.~Li, A.~Cai, L.~Wang, W.~Zhang, and B.~Yan, ``Promising generative
  adversarial network based sinogram inpainting method for ultra-limited-angle
  computed tomography imaging,'' \emph{IEEE Sensors J.}, vol.~19, no.~18, p.
  3941, 2019.

\bibitem{dai2018limited}
X.~Dai, J.~Bai, T.~Liu, and L.~Xie, ``{Limited-view cone-beam CT reconstruction
  based on an adversarial autoencoder network with joint loss},'' \emph{IEEE
  Access}, vol.~7, pp. 7104--7116, 2018.

\bibitem{anirudh2018lose}
R.~Anirudh \emph{et~al.}, ``{Lose the views: Limited angle CT reconstruction
  via implicit sinogram completion},'' in \emph{Proceedings of the IEEE
  Conference on Computer Vision and Pattern Recognition}, 2018, pp. 6343--6352.

\bibitem{xiubin2016limited}
X.~Dai, T.~Liu, D.~Hu, S.~Yan, D.~Shi, and H.~Deng, ``{Limited angle cone-beam
  CT image reconstruction method based on geometric image moment},'' 2016.

\bibitem{feldkamp1984practical}
L.~A. Feldkamp, L.~C. Davis, and J.~W. Kress, ``Practical cone-beam
  algorithm,'' \emph{Josa a}, vol.~1, no.~6, pp. 612--619, 1984.

\bibitem{hammernik2017deep}
K.~Hammernik, T.~W{\"u}rfl, T.~Pock, and A.~Maier, ``A deep learning
  architecture for limited-angle computed tomography reconstruction,'' in
  \emph{Bildverarbeitung f{\"u}r die Medizin 2017}.\hskip 1em plus 0.5em minus
  0.4em\relax Springer, 2017, pp. 92--97.

\bibitem{zhao2018unsupervised}
\BIBentryALTinterwordspacing
J.~Zhao, Z.~Chen, L.~Zhang, and X.~Jin. (2018) Unsupervised learnable sinogram
  inpainting network ({SIN}) for limited angle {CT} reconstruction. [Online].
  Available: \url{https://arxiv.org/abs/1811.03911}
\BIBentrySTDinterwordspacing

\bibitem{zhao2018sparse}
Z.~Zhao, Y.~Sun, and P.~Cong, ``Sparse-view {CT} reconstruction via generative
  adversarial networks,'' in \emph{2018 IEEE Nuclear Science Symposium and
  Medical Imaging Conference Proceedings (NSS/MIC)}.\hskip 1em plus 0.5em minus
  0.4em\relax IEEE, 2018, pp. 1--5.

\bibitem{lee2019high}
D.~Lee, S.~Choi, and H.-J. Kim, ``High quality imaging from sparsely sampled
  computed tomography data with deep learning and wavelet transform in various
  domains,'' \emph{Med. Phys.}, vol.~46, no.~1, pp. 104--115, 2019.

\bibitem{Zhou_2013_ICCV_Workshops}
E.~Zhou, H.~Fan, Z.~Cao, Y.~Jiang, and Q.~Yin, ``Extensive facial landmark
  localization with coarse-to-fine convolutional network cascade,'' in
  \emph{Proceedings of the IEEE International Conference on Computer Vision
  (ICCV) Workshops}, June 2013.

\bibitem{claus2019videnn}
M.~Claus and J.~van Gemert, ``{ViDeNN}: Deep blind video denoising,'' in
  \emph{Proceedings of the IEEE Conference on Computer Vision and Pattern
  Recognition Workshops}, 2019, pp. 0--0.

\bibitem{tassano2020fastdvdnet}
M.~Tassano, J.~Delon, and T.~Veit, ``{FastDVDnet}: Towards real-time deep video
  denoising without flow estimation,'' in \emph{Proceedings of the IEEE/CVF
  Conference on Computer Vision and Pattern Recognition}, 2020, pp. 1354--1363.

\bibitem{armato2010we}
S.~Armato \emph{et~al.}, ``{WE-B-201B-02}: {The} lung image database consortium
  {(LIDC)} and image database resource initiative {(IDRI): A} completed public
  database of {CT} scans for lung nodule analysis,'' \emph{Med. Phys.},
  vol.~37, no. 6Part6, pp. 3416--3417, 2010.

\bibitem{kingma2014adam}
\BIBentryALTinterwordspacing
D.~P. Kingma and J.~Ba. (2014) Adam: A method for stochastic optimization.
  [Online]. Available: \url{https://arxiv.org/abs/1412.6980}
\BIBentrySTDinterwordspacing

\bibitem{maas2013rectifier}
A.~L. Maas, A.~Y. Hannun, and A.~Y. Ng, ``Rectifier nonlinearities improve
  neural network acoustic models,'' in \emph{in ICML Workshop on Deep Learning
  for Audio, Speech and Language Processing}.\hskip 1em plus 0.5em minus
  0.4em\relax Citeseer, 2013.

\bibitem{ledig2017photo}
C.~Ledig \emph{et~al.}, ``Photo-realistic single image super-resolution using a
  generative adversarial network,'' in \emph{2017 IEEE Conference on Computer
  Vision and Pattern Recognition (CVPR)}, 2017, pp. 105--114.

\end{thebibliography}
	
\end{document}